A historical record of four black rainstorm episodes in Hong Kong, China in July–August 2025

P.W. Chan, Y.T. Kwok

Hong Kong Observatory, Hong Kong, China

Corresponding author:

P.W. Chan

Email: pwchan@hko.gov.hk

Hong Kong Observatory, 134A Nathan Road, Kowloon, Hong Kong, China

Abstract

Four episodes of black rainstorms, the highest tier of heavy rain according to the rainstorm warning system in Hong Kong, occurred within eight days from 29 July to 5 August 2025, breaking the record for torrential rain in the territory. Providing early alerts to the public about these black rainstorms has considerable application value and has a substantial impact on their lives. This initial review of the four black rainstorm cases focuses on the predictability of the events from the synoptic/mesoscale to the micro/storm scale. A newly available tool for three-dimensional wind fields retrieved from weather radars in the region was used to analyse the triggering factors for the development and maintenance of intense convection. Given the current level of technology, the four episodes of heavy rain had different degrees of predictability. Global artificial intelligence (AI) models were found to be skilful in capturing two major rainstorms, including one week before they occurred. Surface convergence, mid-tropospheric waves, and jets were important mechanisms in heavy rainfall occurrence. Future research directions on these black rainstorms have been briefly discussed.

Introduction

Located on the southern coast of China, Hong Kong experiences heavy rain in the summer, which is associated with the southwest monsoon and tropical cyclones. The Hong Kong Observatory (HKO) is a meteorological authority in Hong Kong providing weather forecasts and warnings to the public and special users. In relation to rain, a rainstorm warning system (https://www.hko.gov.hk/en/wservice/warning/rainstor.htm) is operated as a three-tier system, namely, amber, red, and black, when heavy rain has fallen or is expected to fall over Hong Kong, exceeding 30/50/70 mm in an hour, and is likely to continue.

Since the implementation of the current system in 1998, there have been at most three episodes of black rainstorm warnings in 2000 and 2006. 2025 will be the first time that there has been four episodes of black rainstorm warnings in one year. The duration of the black rainstorm warning on 5 August 2025 was nearly 11 h, the second longest on record. The longest was the historical rainstorm in September 2023 (Tam et al. 2025). The daily rainfall on 5 August 2025 at the HKO headquarters reached 368.9 mm, that is, the highest daily rainfall in August since records began in 1884. The four episodes of black rainstorms occurred within eight days. They are considered to be historical events that require documentation for future reference.

In this study, we have documented the synoptic and mesoscale features of four episodes of black rainstorms in Hong Kong in July and August 2025. This study is novel in two respects: (1) the microscale/storm-scale features of rainstorms are analysed based on three-dimensional wind fields retrieved from weather radars (Chan et al., (2025)), and (2) the warning aspects of rainstorms are discussed based on the forecasting aspects and predictability of events. A more detailed analysis and prediction of the evolution of heavy rain events are currently being conducted and will not be covered in the present paper.

Synoptic situation

The surface isobaric charts for the four days with black rainstorm warnings are shown in Figure 1. During the entire period, an extensive area of low pressure covered continental China with relatively slackened surface isobars over the southern coast of China. In the first black rainstorm case (29 July 2025; Figure 1(a)), tropical cyclone Co-may affected the East China Sea. The associated low pressure covered the entire southern part of China. A similar situation persisted for the second black rainstorm (2 August 2025; Figure 1(b)). However, the low pressure area near Beibu Wan helped bring a slightly stronger south-westerly wind to the coast of Guangdong, including Hong Kong. For the two remaining cases on 4 and 5 August 2025 (Figures 1(c) and 1(d), a typical southwest monsoon pattern was observed over southern China. Moderate south-westerly to south-westerly winds prevailed over this region. As shown in the surface isobaric charts, surface forcing was relatively weak, and the pressure patterns alone did not appear to be conducive to heavy rain over Hong Kong. They showed that the southwest monsoon could be active in the last two cases (4 August and 5, 2025). From the synoptic analogue based on the European Centre of Medium Range Weather Forecast (ECMWF) reanalysis, no past rainstorm cases could be identified for the first and second days. At most, a couple of past red rainstorm cases were identified on the third and fourth days.

The 850 hPa wind and humidity fields (Figure 2) and 500 hPa geopotential height (Figure 3) show the flow patterns in the lower and middle troposphere. They are based on a real-time analysis of the ECMWF Integrated Forecasting System (IFS). In the first case, a slight troughing flow with enhanced vorticity was observed near the southeastern coast of China at 850 hPa (Figure 2(a)) and 500 hPa (Figure 3(a)). However, the vorticity did not seem to be particularly high in triggering rainstorms over the coast of Guangdong. In the second case, a south-westerly jet was observed at 850 hPa along the southern coast of China (Figure 2(b)). However, the winds were not particularly strong at mid-level (Figure 3(b)). For the third and fourth cases, there were typical patterns of an active southwest monsoon in the lower troposphere (Figures 2(c) and 2(d)). The extensive vortex in the middle troposphere was indicative of a chance of heavy rain (Figures 3(c) and 3(d)). As such, the synoptic to mesoscale forcing for the first and second rainstorms was relatively weak and did not seem to suggest black rainstorms over Hong Kong. However, the lower to middle tropospheric patterns in the synoptic to mesoscale range suggest that heavy rain may be expected over southern China. This is consistent with the forecasts of the global numerical weather prediction (NWP) models and global artificial intelligence (AI)-based models discussed in a later section.

In the upper troposphere (200 hPa, Figure 4), a substantial divergence was observed near Hong Kong in all cases. However, divergence near the territory was isolated for the first two black rainstorms (Figures 4(a) and 4(b)), with the respective speed and directional divergence embedded within a board anticyclone. Accurate forecasting of divergence is rather difficult, and a slight shift in the divergence location would indicate completely different rainfall situations over Hong Kong. In contrast, extensive and significant divergence was observed for the third and fourth black rainstorms (Figures 4(c) and 4(d)) in association with an easterly wave near Taiwan, which may have a higher predictability.

Based on synoptic and mesoscale analyses, the chance of having a rainstorm over Hong Kong was higher for 4 August and 5, 2025, based on the larger-scale flow patterns, and thus a higher predictability of the events. It is still difficult to ascertain whether there could be torrential rain over the territory because of the need to forecast the micro-to storm-scale features. However, for the 29 July and 2 August cases, the hints from synoptic and mesoscale patterns for heavy rain occurrence are substantially lower. As shown in the section below, the torrential rain over Hong Kong is mostly related to small-scale features just over the coast of Guangdong. The predictability of these two black rainstorm events, even one day ahead, is rather limited, given current technology.

Dynamical and thermodynamical parameters

The 00 UTC and 12 UTC radiosonde ascents in Hong Kong were analysed daily based on several dynamic, thermodynamic, and stability-related parameters. The values were compared with the climatology of Hong Kong over the last 20 years in the same period (July to September). The 00 UTC analysis of the four rainstorm cases under consideration is shown in Figure 5.

The climatological ranges of these values are shown as box-and-whisker plots. The current 00 UTC radiosonde values are shown as broken horizontal lines. If this value is exceptional compared to the climatology, the corresponding box for that value is highlighted in red (exceptional) or purple (extreme). The global NWP model predictions at that time are indicated by arrows on the right side of each box.

A common feature of the four rainstorm cases was the relatively high total precipitable water vapour (PWV) value, in excess of 70 mm. Based on an analysis of the historical rainstorms of 2023 (Tam et al., 2025), this value is conducive to the occurrence of rainstorms. From the measurements of ground-based microwave radiometers, the PWV value once reached 80.4 mm at 15 UTC on 4 August 2025 during the third episode of black rainstorm and 86.6 mm at 03 UTC on 5 August 2025 during the fourth episode of black rainstorm. These two are exceptionally high values based on climatology of Hong Kong.

Apart from that, for the first two rainstorm cases (Figures 5(a) and 5(b)), the dynamic and stability-related parameters are not particularly exceptional in terms of climatology, apart from a K-index value of nearly 40 °C. The humidity-related parameters were generally high (highlighted in red and purple boxes). These features show that heavy rain may have occurred on these two days. However, it is not clear whether the rain was particularly heavy enough to cause black rainstorms in Hong Kong.

However, for the third and fourth cases (Figures 5(c) and 5(d)), the dynamic parameters (first column) and humidity-related parameters (second column) were exceptional when compared with the climatology (all boxes highlighted in red). The vertical wind shear supports the development of intense convection. For stability-related parameter, K-index also reaches rather high values of approximately 40 °C. Coupled with synoptic and mesoscale features, troposphere conditions also suggest heavy rain over Hong

Kong. Therefore, the background atmospheric conditions show that torrential rain is possible over the territory. The remaining challenge is whether convection is so organised that black rainstorms are expected over Hong Kong. In this region, the operation of rainstorm warnings is based on nowcasting.

Mesoscale/microscale features based on weather radars and surface observations

For the first time on 29 July 2025, a surface convergence line was identified in the Pearl River Estuary before and during rain events (Figure 6(a)). This is related to (1) weak northerly winds in the board low circulation and possible enhancement in the nighttime due to land breezes and (2) setting in of south to south-westerly winds in the early morning due to the southwest monsoon in the summer (Li et al., 2018). In this region, the upward motion was found to be quite extensive and persistent from the vertical velocity field at a height of 2 km above sea level, as analysed from the weather radar Doppler velocities (Figures 6(b) and 6(c)). The updraft is strong and persists for a long time, reaching a height of at least 10 km above the sea surface, where the northerly outflow in the upper troposphere closes the sustained vertical circulation. Surface convergence and vertical circulation appear to assist in the development of intense convection. This persists for a prolonged period over Hong Kong without substantial storm motion due to relatively weak mid-tropospheric winds.

The surface convergence situation was relatively similar in the second case (2 August 2025; Figure 7(a)). However, the wind direction was more variable in the inland area without a dominant wind direction, as in the first case. The surface wind pattern was similar to that of a typical southwest monsoon in the early morning. However, intense convection develops and is maintained over the coast in the Pearl River Estuary area in the form of an east–west oriented rainband. From the three-dimensional wind field retrieved from the weather radar, there appears to be a chain of upward and downward motions in the vertical velocity field at a height of 2 km or above, such as the 2 km vertical velocity field in Figure 7(b). When cross sections are made of radar echoes (Figures 7(c) and 7(d)), there are substantial waves in the middle to upper troposphere with strong upward and downward motions. Waves appear between 2 and 9 km above sea level (ASL). The storm-scale feature, namely, persistent perturbations of the middle to upper tropospheric westerlies to southwesterlies, may be a major mechanism for sustaining the east–west oriented rainband.

For the third and fourth cases on 4 and 5 August, 2025, a strong south-westerly flow was found at a height of 3 km ASL in the constant altitude plan position indicator (CAPPI) scans of the weather radar in Hong Kong. Within this strong airflow, there appeared to be a number of jet streaks, labelled A to E, occurring occasionally (left panels of Figures 8(a)–8(e)) with a repeating cyclone of approximately 12 min. In association with this, stronger radar echoes moved from west to east (right panels of Figures 8(a)–8(e)). As in the second black rainstorm case, perturbations could be identified in the strong west to south-westerly airflow at a height of 3 km above sea level, as analysed in the three-dimensional wind field. The occurrence and periodicity of jet streaks (and the associated strong radar echoes) require further investigation, for example, using storm-scale simulations. However, the repeated occurrence of jet streaks assists in the development and persistence of a generally east-west oriented airflow.

As in the first black rainstorm case, the vertical circulation of the rainband was studied further using a vertical cross section, as shown in Figure 9. A substantial updraft was observed at the location of intense radar echoes, and closed-loop circulation was found in the flow across the prevailing south to south-westerly. This vertical circulation helps maintain the convection and development of heavy rain. However, its occurrence requires further investigation.

Stronger winds occur in the middle of the troposphere. Bursts of high winds are also recorded at the surface from time to time, adding to the rather poor perception of weather conditions. An example wind burst was captured using the Doppler velocity of the weather radar, as shown in Figure 10(a). A vertical cross section of the weather radar echo was used to study the wind field, as shown in Figure 10(b). Behind the squall-line-like feature in the weather picture, there is an extensive area of high winds, with wind speeds reaching storm force at approximately 50 knots (coloured purple in Figure 10(b)) and descending motion (at least down to a height of 2 km ASL). This descending jet appears to be separate from the radar echo in the front, that is, it does not appear to be a typical rear-inflow jet of a squall line. The occurrence of jet streaks requires further investigation, for example storm-scale simulations.

For the third and fourth black rainstorms, some storm-scale features were noticeable for developing and maintaining intense convection, such as mid-tropospheric jet streaks and cross-flow vertical circulation. Jet streaks may reach the ground, leading to bursts of strong winds experienced by people on the surface. The alignment of the timing of mid-tropospheric jets and surface wind gusts is under investigation and will be reported in a future paper focusing more on storm-scale dynamics.

Forecasting aspect

As an example, the global AI model Pangu based on ECMWF analysis, that is, the so-called Pangu-EC was studied (He and Chan, 2025), using a model run initialised at 00 UTC on 27 July 2025. Only the forecast surface isobaric patterns and the 500 hPa geopotential height field are shown in Figure 1.

Two days before the first black rainstorm on 29 July 2025 the synoptic and mesoscale patterns at the surface and 500 hPa were effectively captured (Figures 11(a) and 11(b)). However, because of the slight difference in the forecast location of the upper-air divergence compared to the actual observations, a substantial heavy rain area was predicted to the south and southeast of Hong Kong (Figure 11(a), with rainfall highlighted in green/yellow/red). Over the territory, less than 10 mm of rainfall was forecasted. A similar prediction applies to the second black rainstorm of 2 August (Figures 11(c) and 11(d)), namely, a T+144 h (6 d ahead) forecast. The model captures the generally east–west oriented rainband near the southern coast of China. However, the forecast rainband is mainly to the southeast of Hong Kong over the sea. In contrast, the forecast rainfall amount of the territory was relatively small, again less than 10 mm. This may be related to a slight anticyclonic flow pattern immediately over the territory in the middle troposphere (Figure 11(d)) with a low over Guangxi and another low over eastern China, and an anticyclonic flow in between. The upper-air divergence area was also slightly different from Hong Kong (not shown). For various reasons, not all global NWP and AI models are capable of capturing these two black rainstorms in Hong Kong in advance.

The predictive ability of the latter two cases was considerably higher. For Pangu-EC, an east–west oriented heavy rain band was forecast along the southern coast of China at T + 192 8 d ahead and T + 216 9 d ahead. The active surface southwesterlies (Figures 11(e) and (g)) and the 500 hPa extensive board low (Figures 11(f) and 11(h)) were effectively captured, and thus the heavy rainfall was correctly forecasted. These forecast patterns remain almost the same in the subsequent runs (not shown). Therefore, there is a high confidence for HKO to issue an alert message to the public about the chance of heavy rain on 4 August 5, and 2025 (discussed in the next section).

The predictability of events at the sub-seasonal scale was also considered. The ECMWF 46-d forecast is shown here as an example. The predicted 850 hPa anomalous wind field and the predicted rainfall anomalous field are shown in Figure 12. For the model run initialised at 00 UTC on 13 July 2025 an unseasonally strong south-westerly flow in the lower troposphere was already predicted for southern China for weeks 3 (Figure 12(a)) and 4 (Figure 12(b)). Abnormally high rainfall was forecasted in the former period. but there did not seem to be a strong signal of rainfall in the latter period. One week later, namely, for the model run initialised at 00 UTC on 20 July 2025 the same story remains (Figure 11(c) for week 2 forecast, and Figure 11(d) for week 3 forecast). Similar forecast patterns were observed in other sub-seasonal forecast models used for the HKO (not shown). As such, the sub-seasonal models were not found to be particularly skilful in capturing the occurrence of historical black rainstorm episodes during the 8-d period from 29 July to 5 August 2025. This is likely partly because of the dominant factor of microscale/storm-scale features in determining the location and intensity of heavy rain over such a small place as Hong Kong.

Implications for the operation of rainstorm warning service

Located along the southern coast of China and influenced by the southwest monsoon, Hong Kong normally experiences maximum rainfall in the early morning of summer (Chan and Ng, 1993; Li et al., 2018). To alert the public of the potential occurrence of rainstorms the next day, especially during the early morning hours of school and work, it is highly desirable to provide a clear statement about the situation of rain in the afternoon of the previous day. If this is not possible, the present protocol is to give the alert, at least for schooling, between 5 am and 5:30 am in the morning, about the chance of a rainstorm in the next few hours. However, such an alert on the afternoon of the previous day, or even in the short hours of the morning, may not be possible if the synoptic and mesoscale patterns are not strongly indicative of the occurrence of rainstorms, such as the black rainstorms of 29 July 2025 and 2 August 2025 when the surface forcing is weak. The lower and middle tropospheric patterns are rather common in the southwest monsoon in the summer and the upper tropospheric divergence is isolated and may not be correctly forecasted over Hong Kong. Rainstorms appear to be more closely related to mesoscale to microscale features, such as surface convergence and mid-tropospheric waves. These are not well forecasted by global NWP, global AI, and mesoscale NWP models. Therefore, the predictability of this type of rainstorm is rather low, and its occurrence is likely to attract widespread criticism from the public for not providing advance warnings, particularly when the rainstorm occurs during the peak hours of returning to work between 8 am and 9 am. To handle this type of rainstorm, efforts have been made to further strengthen the nowcasting system, for example, based on storm-scale modelling using weather radar data assimilation and AI-based nowcasting methods. These results will be reported in a future study.

In contrast, for the black rainstorms of 4 August and 5, 2025, the heavy rain formation mechanism was mostly related to synoptic and mesoscale flow features. They were captured well by the global NWP models and global AI models, even a week ahead. With proper training, the global AI models persistently gave an east–west oriented rainband along the southern coast of China on those two days. For the Pangu-EC model, since the model run initialised at 00 UTC on 27 July 2025 the maximum daily rainfall in the region of 200 mm to 300 mm has been persistently forecasted at the grid points near Hong Kong on 4 August and 5, 2025. This maximum daily rainfall was consistent with the order of magnitude of the actual observed rainfall. The persistence of forecasting this exceptionally high amount of rainfall gives weather authorities the confidence to give the public an early chance of torrential rain on these two days. In the evening of 4 August 2025 a specific message was issued to the public about particularly heavy rainfall, at first on 5 August 2025. This turned out to be correct and was well received by the public. Although correct in forecasting the east-west oriented rainfall along the southern coast of China, the global NWP models are not able to predict the correct order of magnitude of the maximum daily rainfall. For the ECMWF IFS, the maximum daily rainfall at the grid points near Hong Kong remained below 100 mm, mostly in the order of 50–70 mm.

Regarding the sub-seasonal aspect, the global models captured the unseasonally strong southwest monsoon during this period well. However, potentially owing to limitations in spatial resolution and cloud physics, there is no clear signature of the historical rainstorm on 4 August 5, and 2025, for example, based on the forecast rainfall anomaly. Although such models provide some hints regarding unsettled weather, it is still not possible to highlight historical rainstorms a couple of weeks ahead.

Conclusions

Black rainstorms have a substantial impact on Hong Kong's daily lives because members of the public are advised to stay indoors during the warning period. This affects people commuting to work and school. An earlier indication of the chance of a black rainstorm is highly desirable to prepare people for worse weather. This study has provided an initial and brief review of the predictability of the historical occurrence of four episodes of black rainstorms in Hong Kong over eight days. The four episodes have different degrees of predictability. The cases of 29 July and 2 August appear to be more related to the storm-scale features of Hong Kong and limited predictability days ahead. Meanwhile, the cases of 4 and 5 August have persistent signatures, and advanced public warnings could be possible in the evening before the morning maximum rainfall on the following day. Synoptic, mesoscale, and storm-scale features were reviewed in this study, and the forecasting skills of a global AI model and a sub-seasonal model were discussed for reference in the future.

For a rather small place, such as Hong Kong, the chance of heavy rain is strongly related to microscale and storm-scale features, which seem to have limited predictability beyond the nowcasting scale, that is, a couple of hours before heavy rain occurrence. Further studies are being conducted to reconstruct these four heavy rain events and to explore the temporal scale of nowcasting. It is also being determined whether it is possible to provide an even earlier warning to the public that black rainstorm warning, which has the quantitative requirement of 70 mm of rainfall in an hour, to be widespread in Hong Kong, is going to happen. This can be conducted, for example, using storm-scale forecast by three-dimensional variational analysis (3DVAR) of weather radar data assimilation into a regional NWP model, or a data-driven machine learning model based on past weather radar data for heavy rain cases. These results will have a considerable impact on the operation of rainstorm warning systems in Hong Kong. The results of future studies will be reported in the upcoming papers.

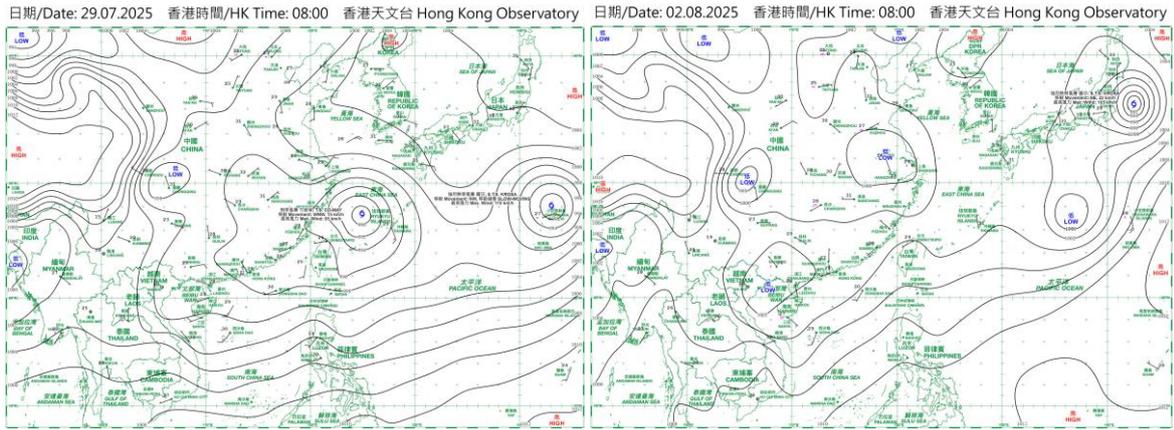

(a) and (b).

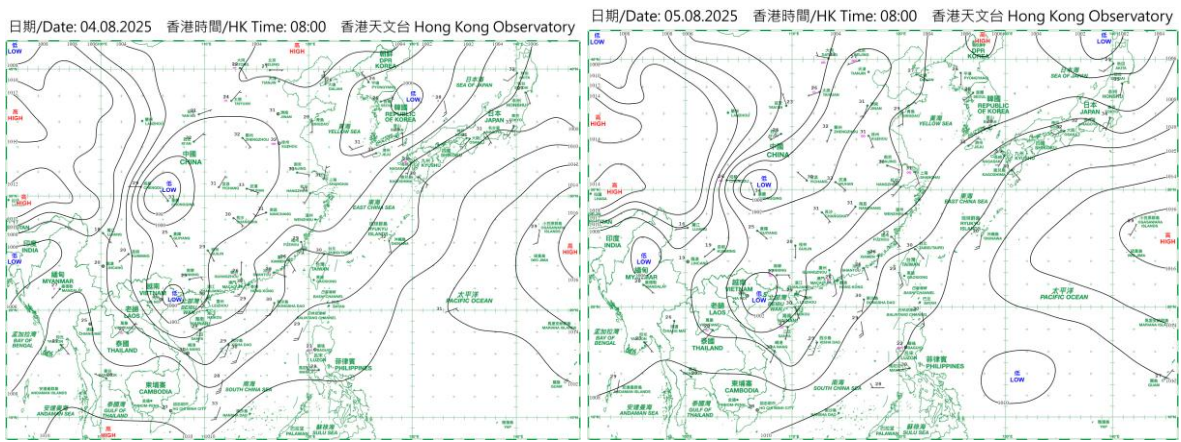

(c) (d)

Figure 1 Surface isobaric charts for the four rainstorm cases – all at 00 UTC (8 am local time, with Hong Kong time = UTC + 8 h) on the respective days.

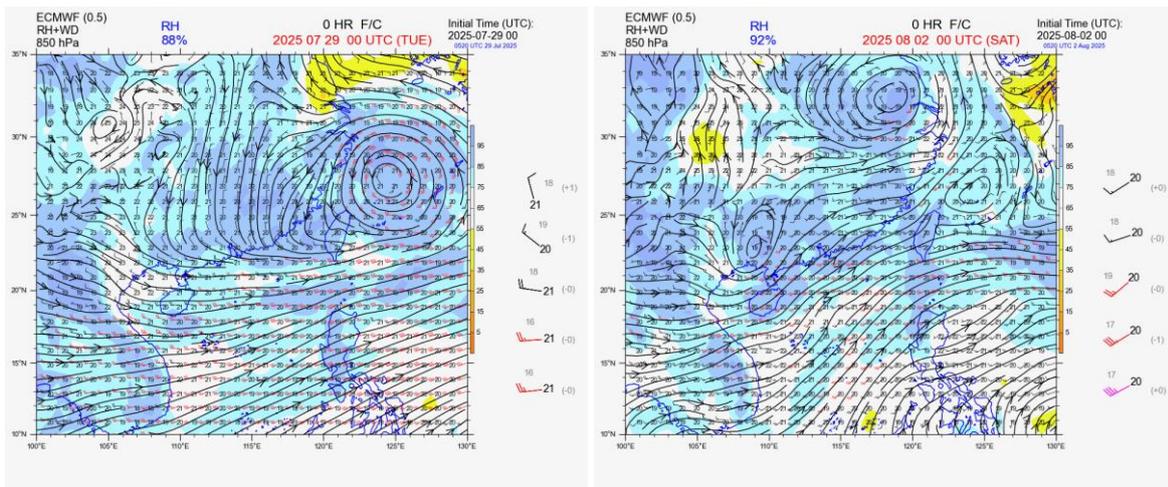

(a) and (b).

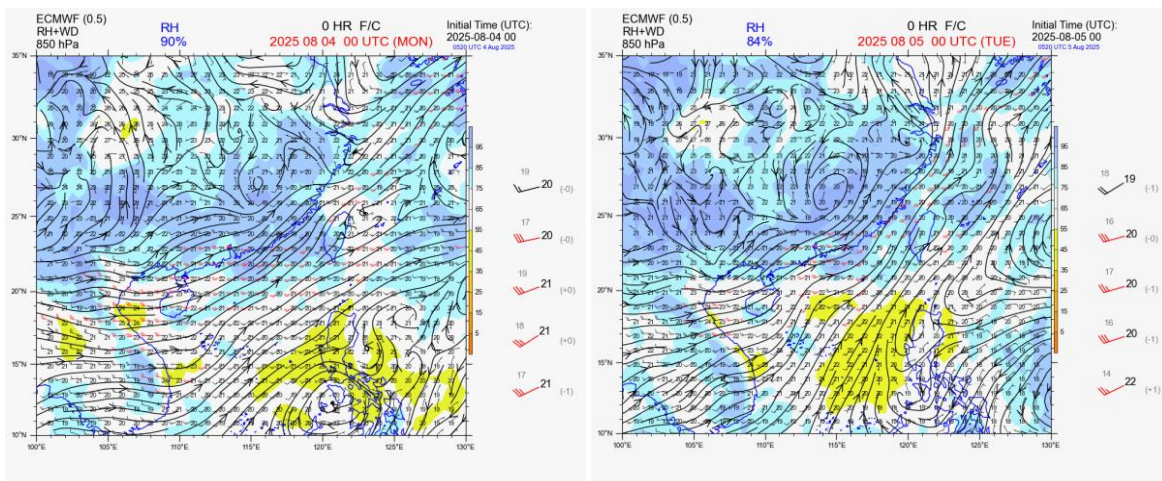

(c) and (d)

Figure 2 The real-time analysis of 850 hPa wind and relative humidity from ECMWF IFS for the four rainstorm cases.

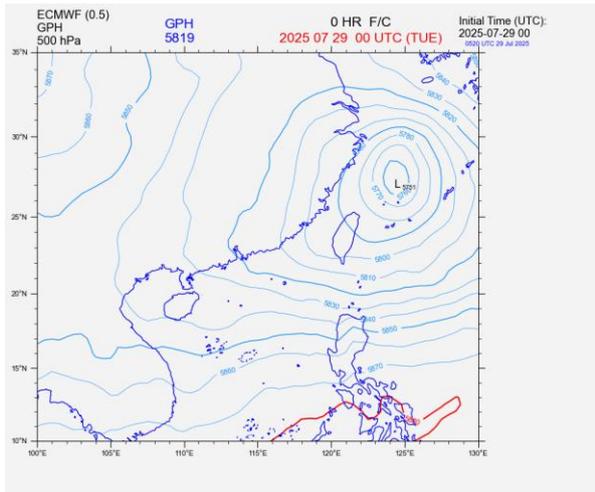 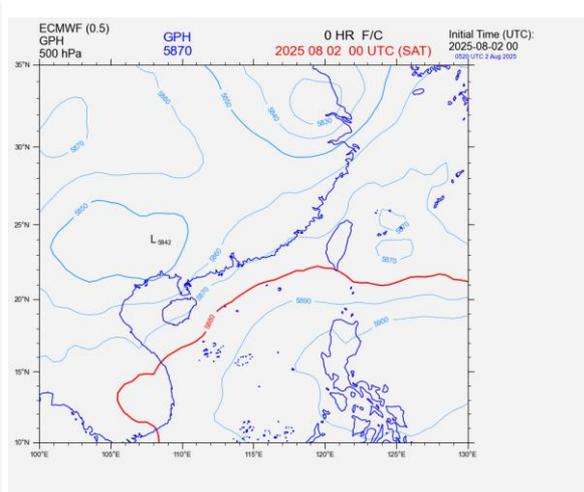

(a) and (b).

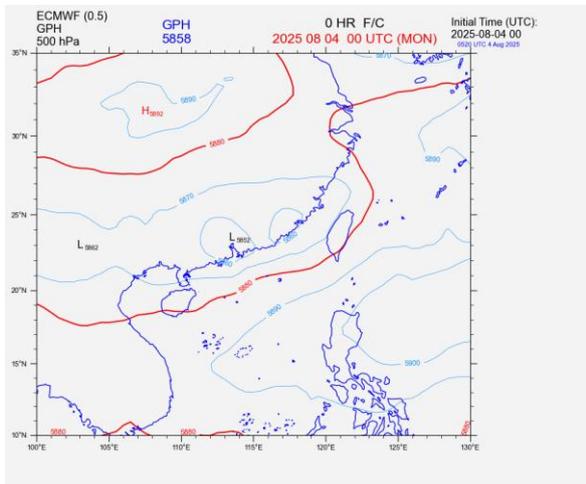 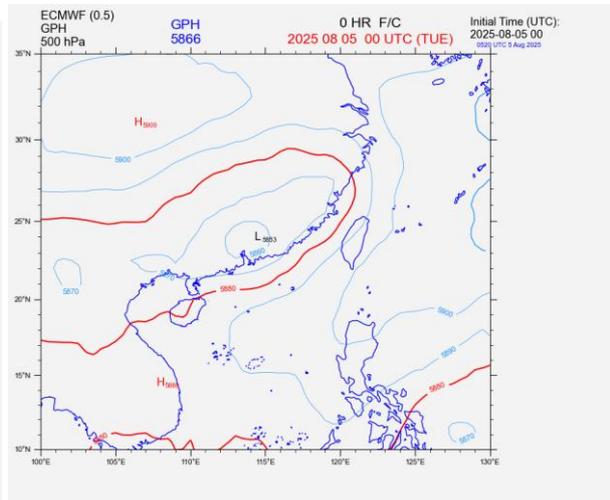

(c) (d)

Figure 3 The real-time analysis of 500 hPa geopotential height from ECMWF IFS for the four rainstorm cases.

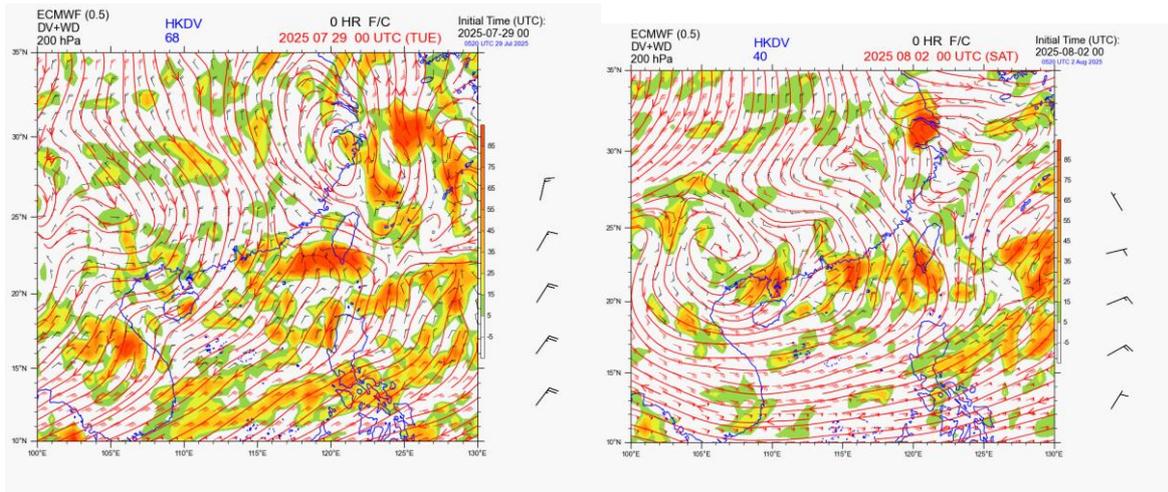

(a) and (b)

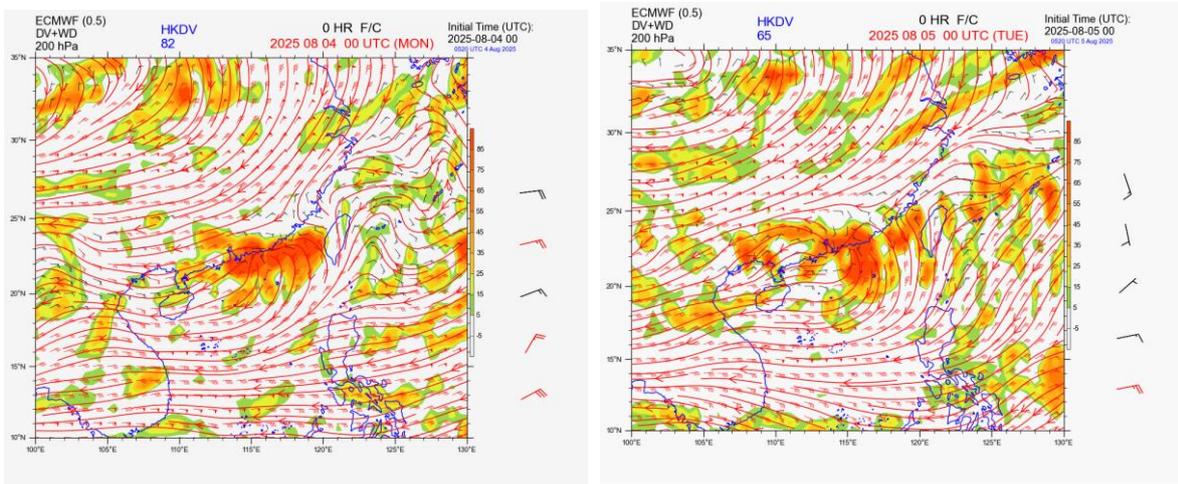

(c) (d)

Figure 4 The real-time analysis of 200 hPa wind and divergence field from ECMWF IFS for the four rainstorm cases.

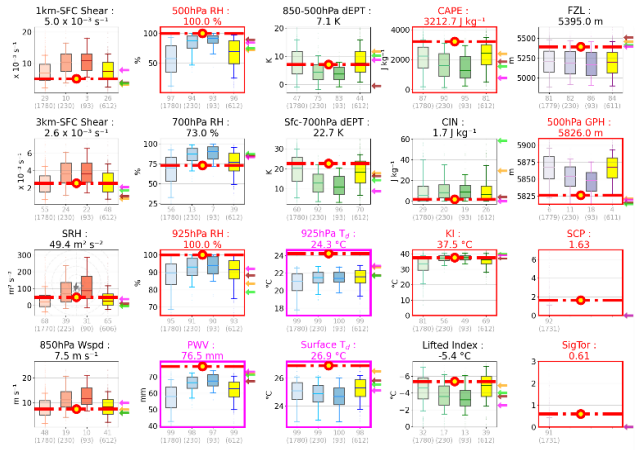
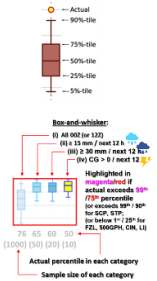
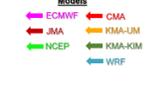
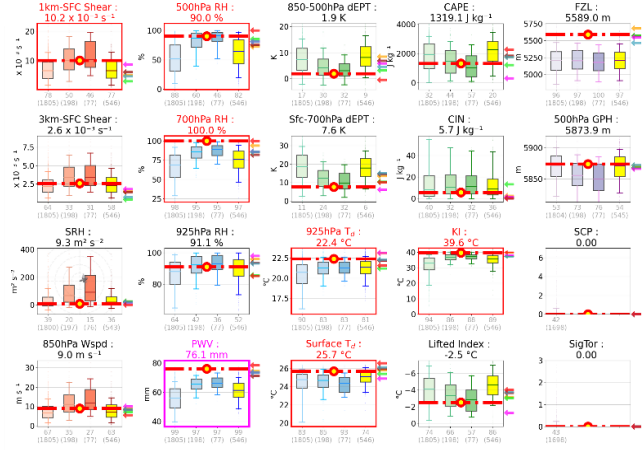
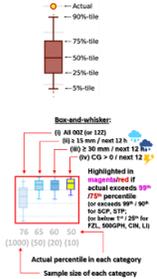
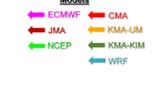

(a) and (b).

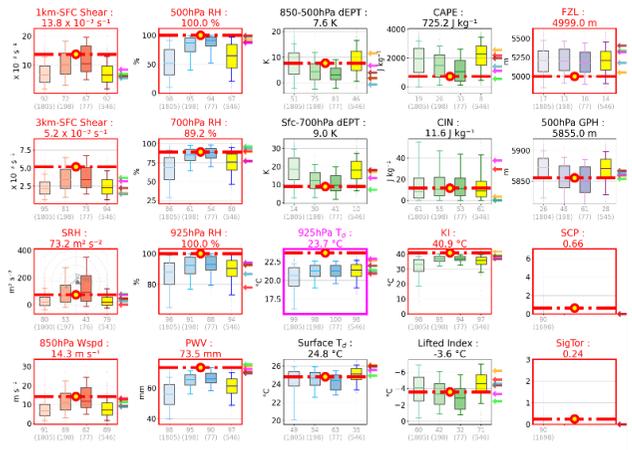
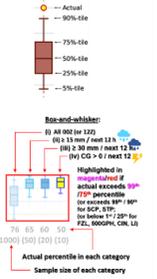
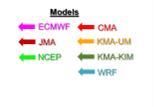
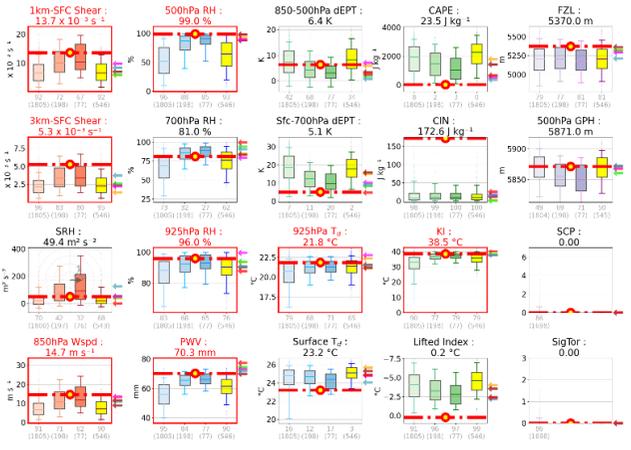
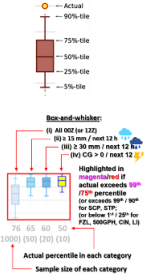
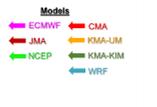

(c) (d)

Figure 5 00 UTC sounding analysis for Hong Kong for the four days with black rainstorms.

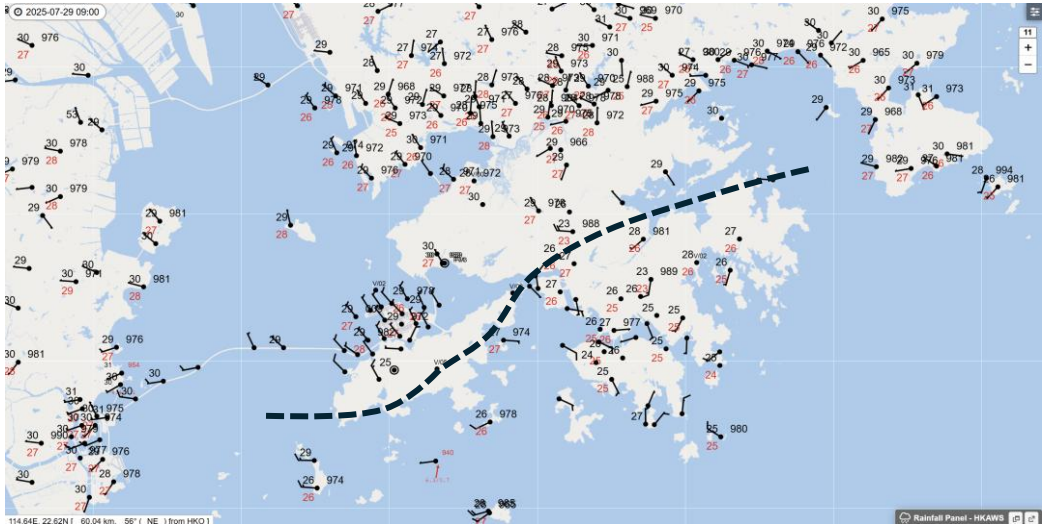

(a)

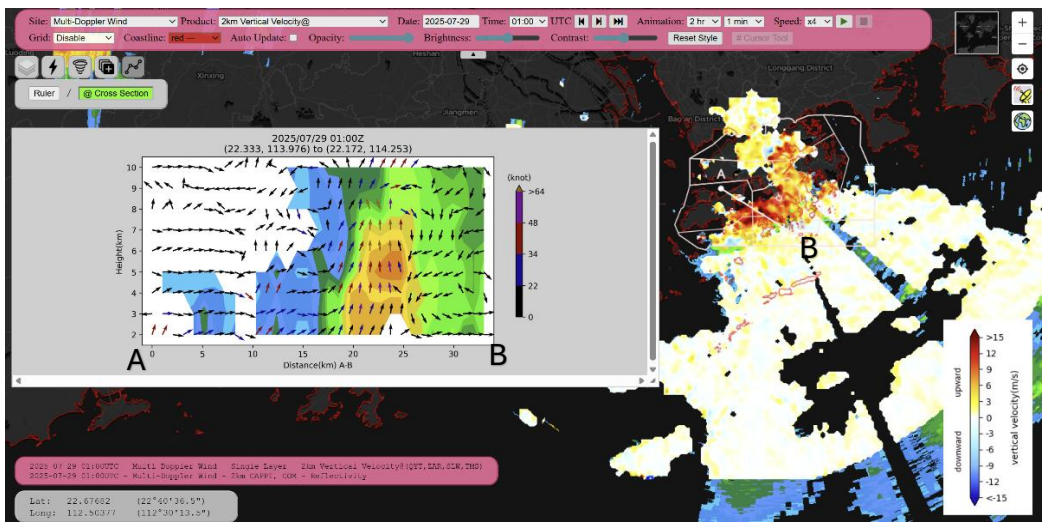

(b)

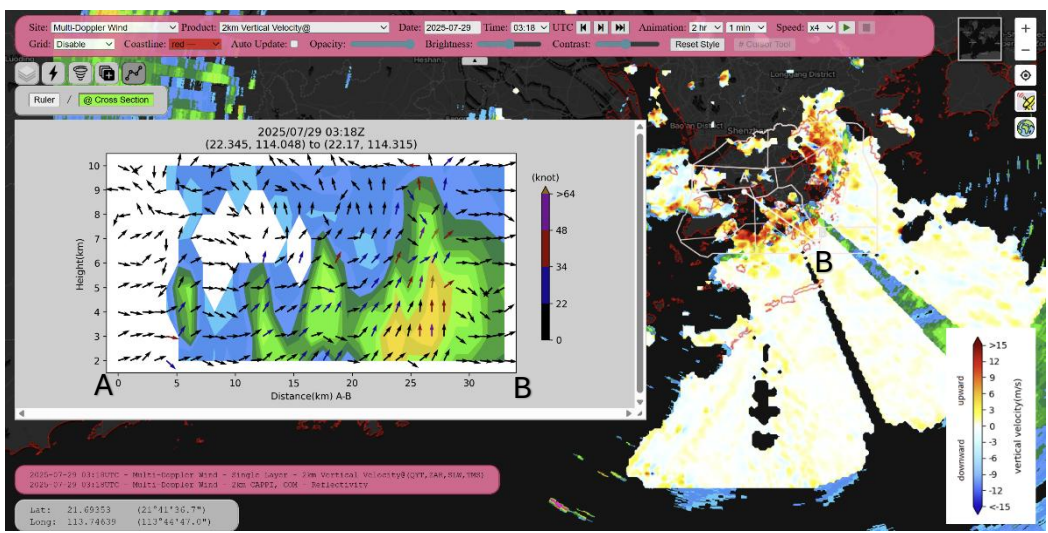

(c)

Figure 6 (a) shows the surface observations in the morning of the first rainstorm case, with the surface convergence highlighted by a broken curve. (b) and (c) are vertical cross sections across the area of the major rainband and the vertical velocity field at a height of 2 km above sea level.

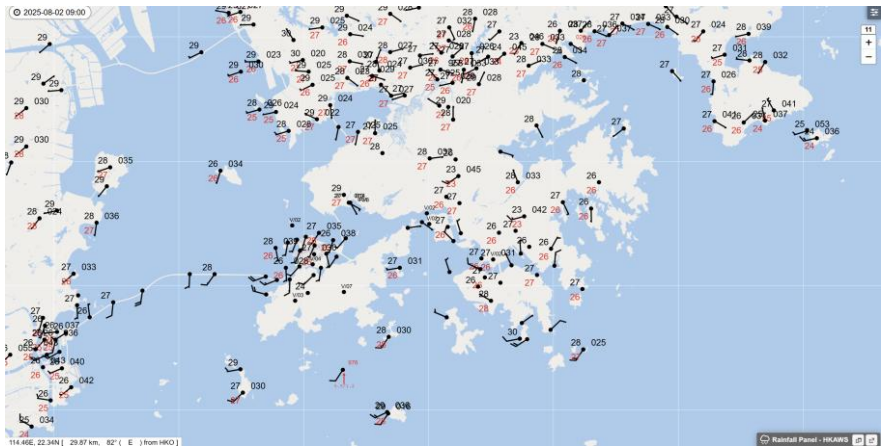

(a)

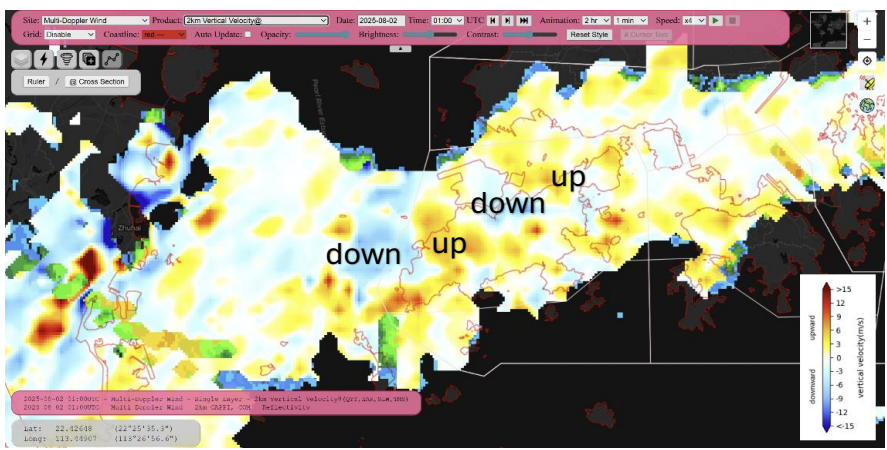

(b)

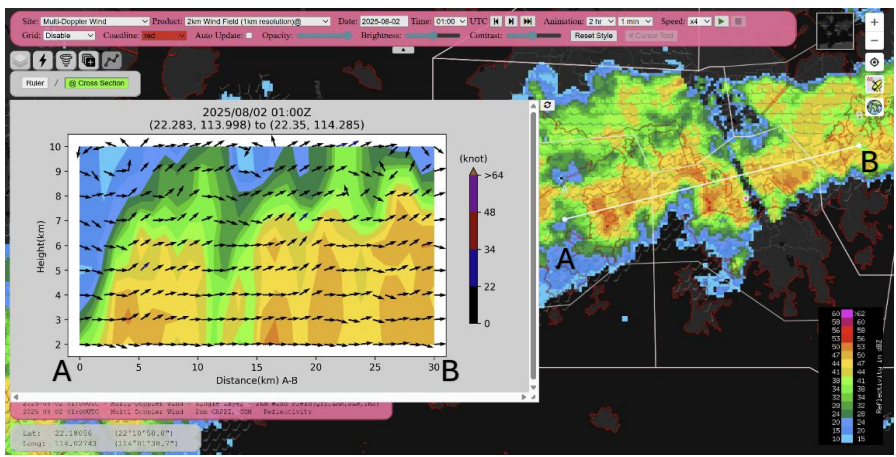

(c)

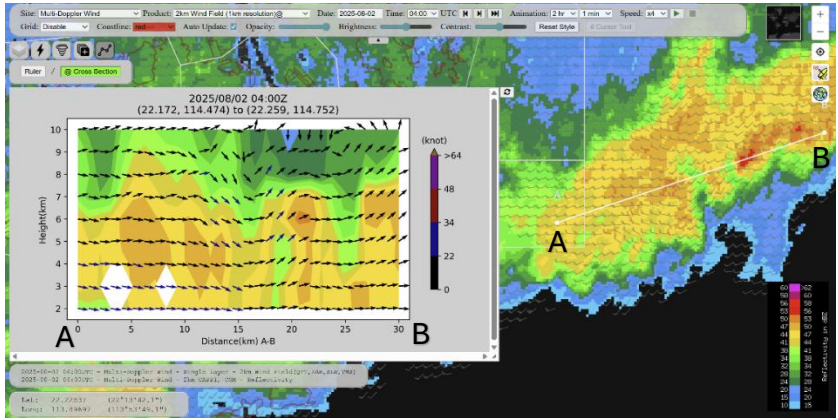

(d)

Figure 7 (a) shows the surface observation of the second rainstorm case; (b) shows the vertical velocity field at a height of 2 km above the sea surface. (c) and (d) show the 2 km wind field overlaid on the weather radar echoes and vertical cross sections.

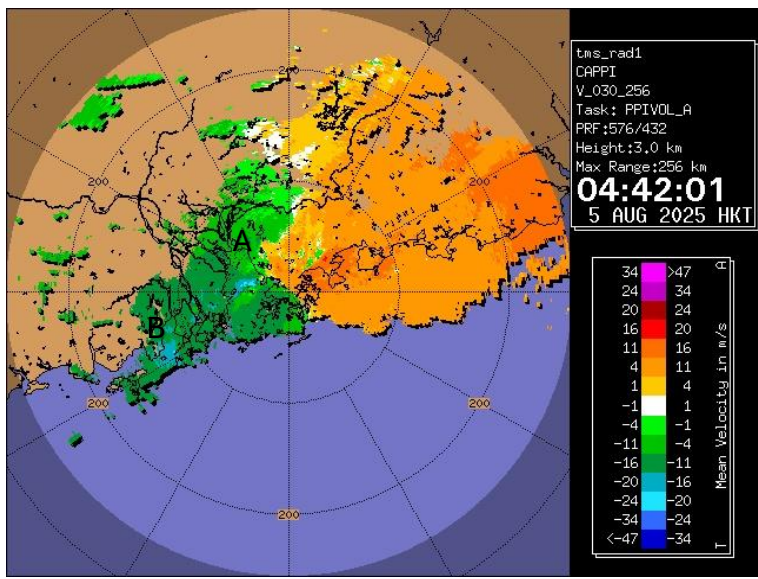
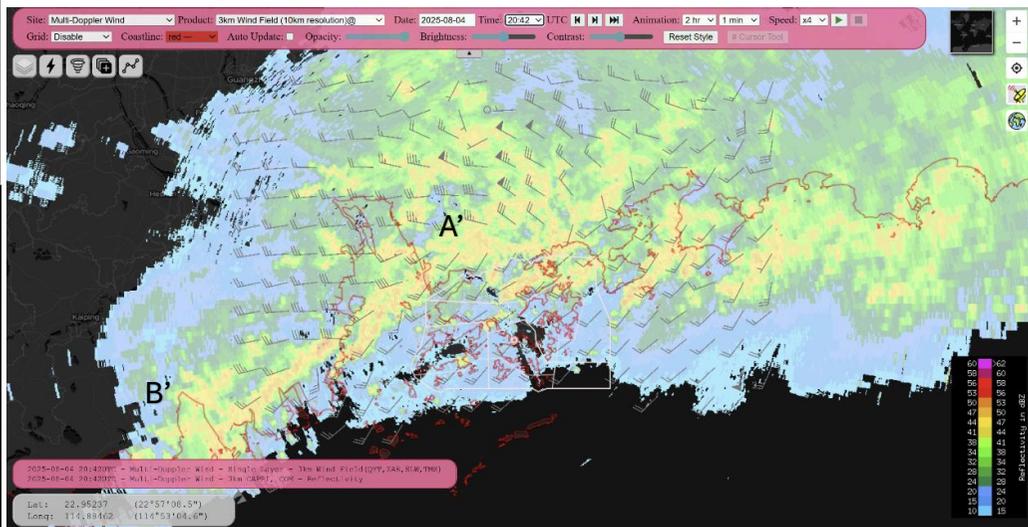

(a)

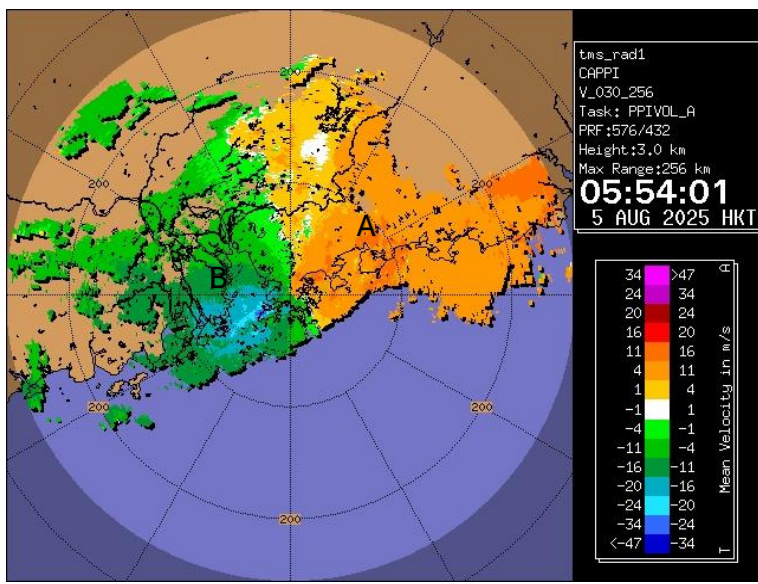
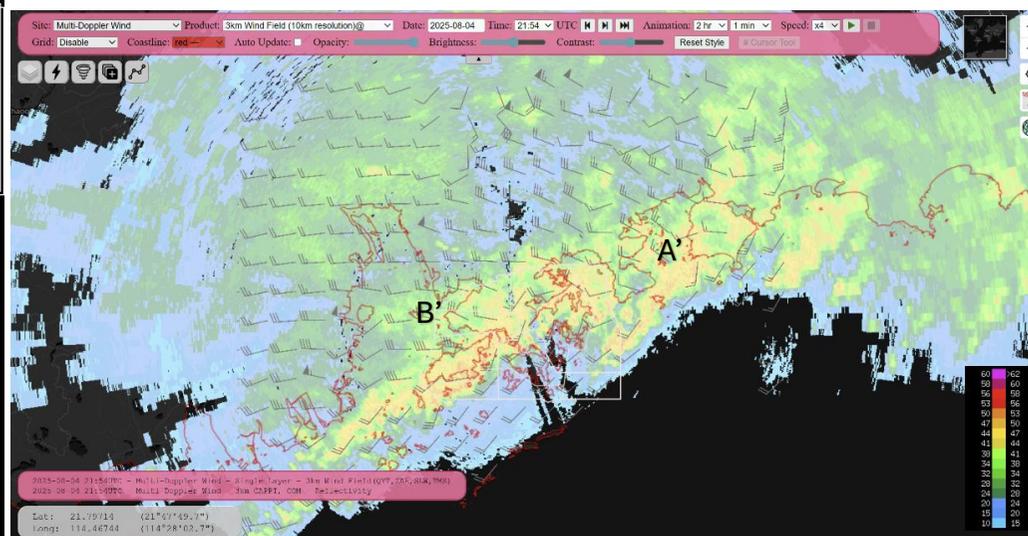

(b)

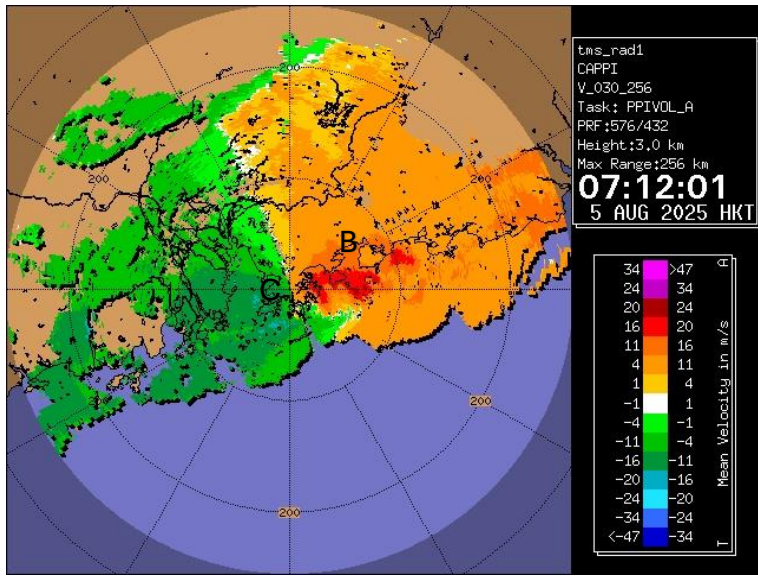
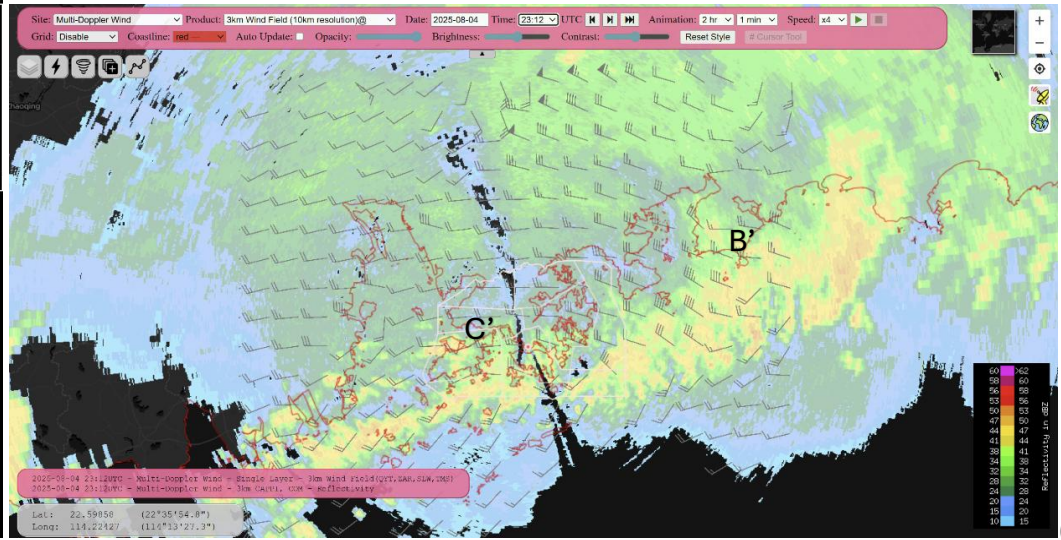
(c)

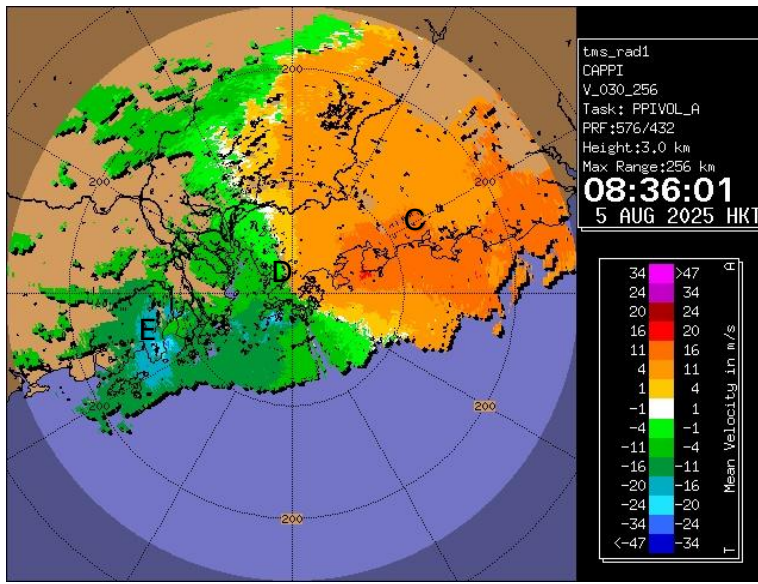
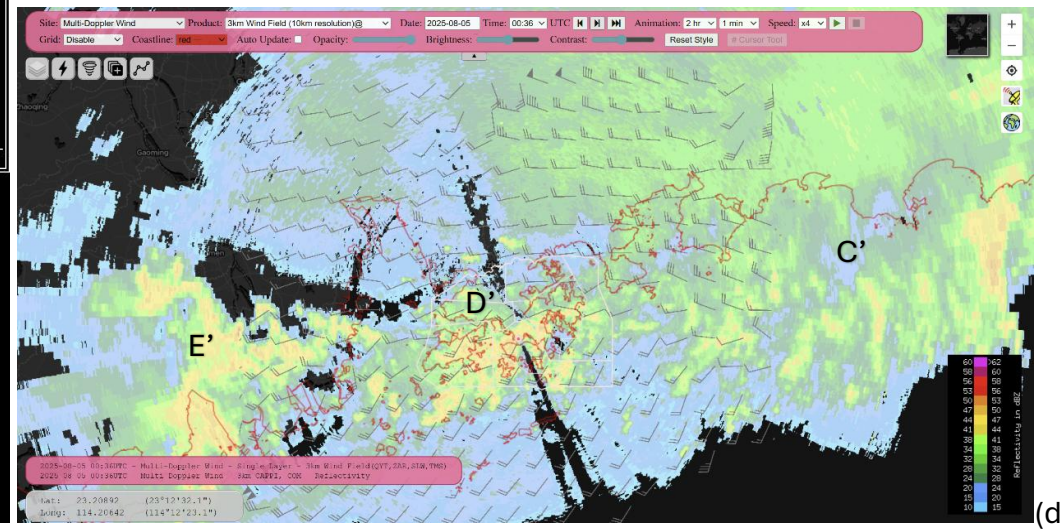
(d)

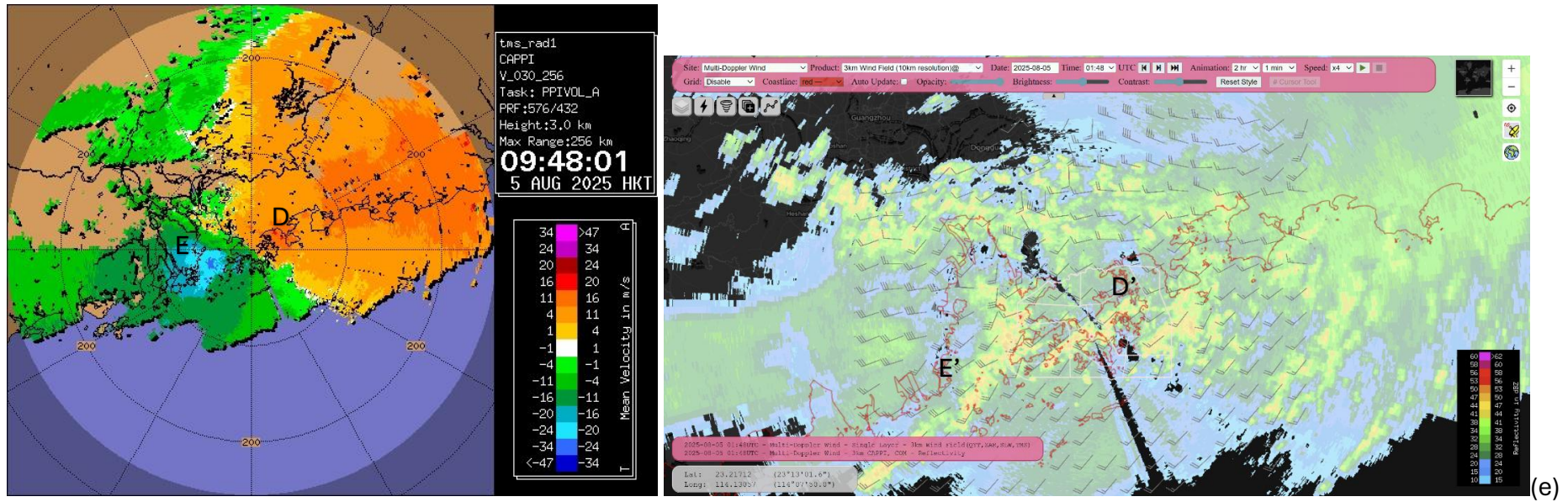

Figure 8 3 km CAPPI Doppler velocity field (left panels) from a weather radar in Hong Kong and the retrieved wind field overlaid with radar echoes (right panels). The wind streaks are labelled by A to E on the right panel and the associated radar echoes are labelled A' to E' on the right panels.

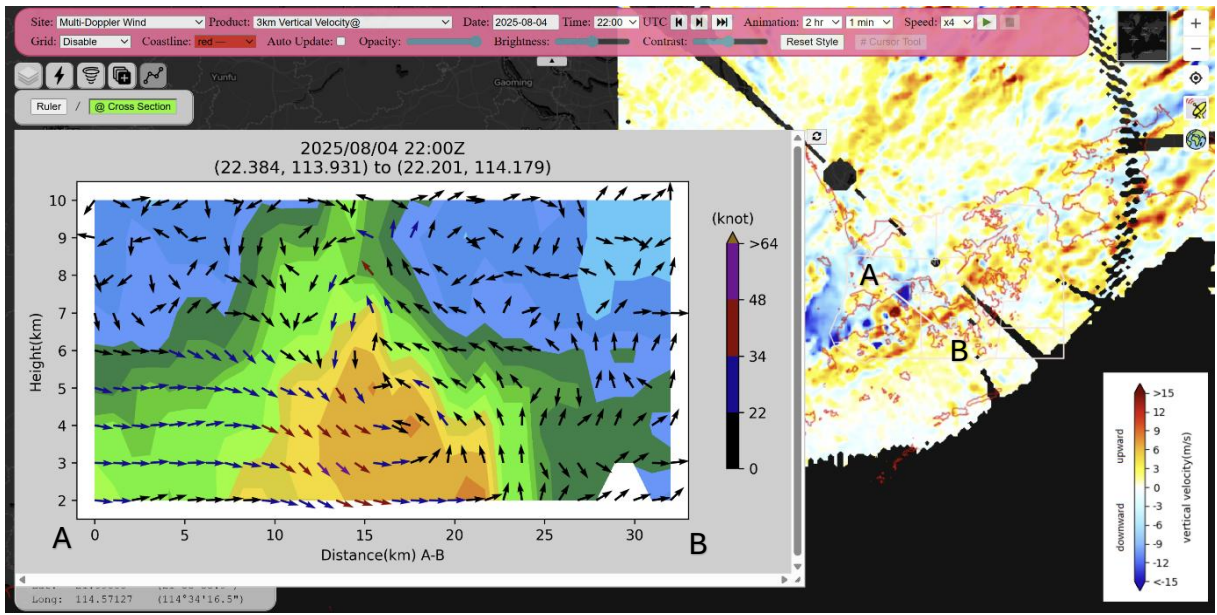

Figure 9 A snapshot of the 3 km vertical velocity field for the fourth rainstorm case and a vertical cross section showing the vertical circulation.

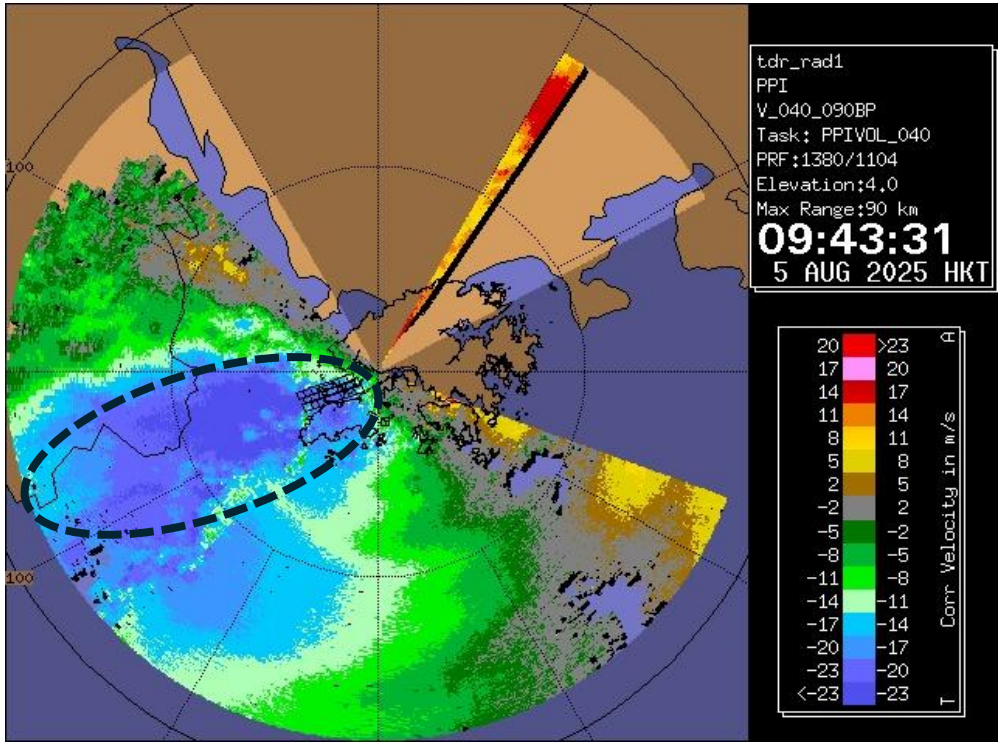

(a)

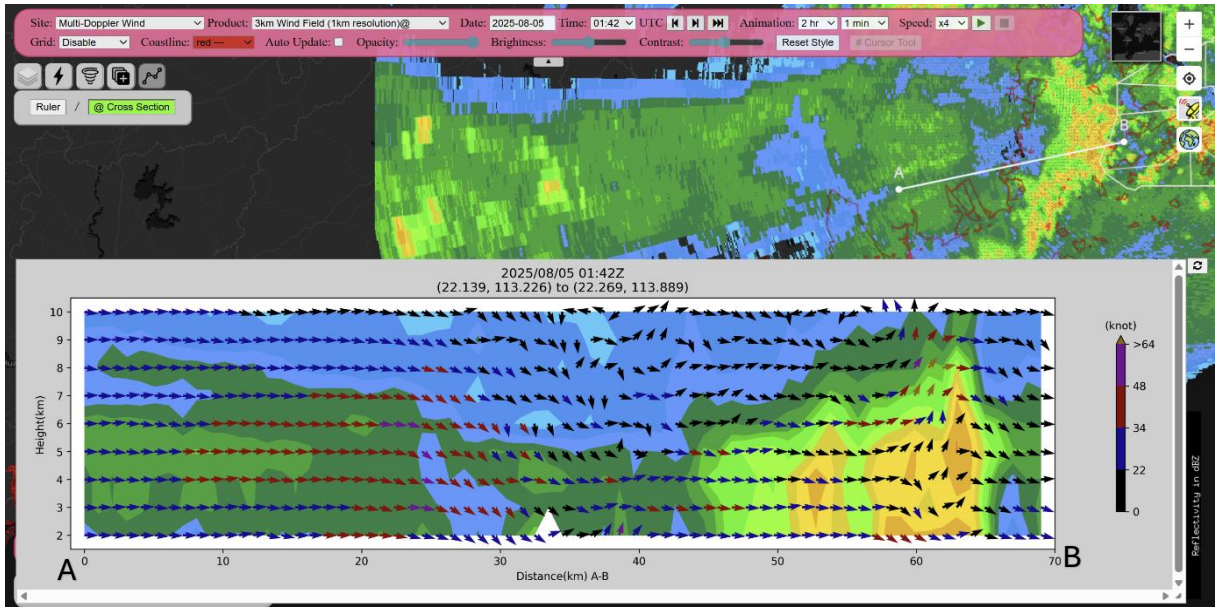

(b)

Figure 10 Doppler velocity field from a weather radar in Hong Kong with the wind burst highlighted in an ellipse (a) and the 3 km wind field as well as the vertical cross section in the jet streak (b).

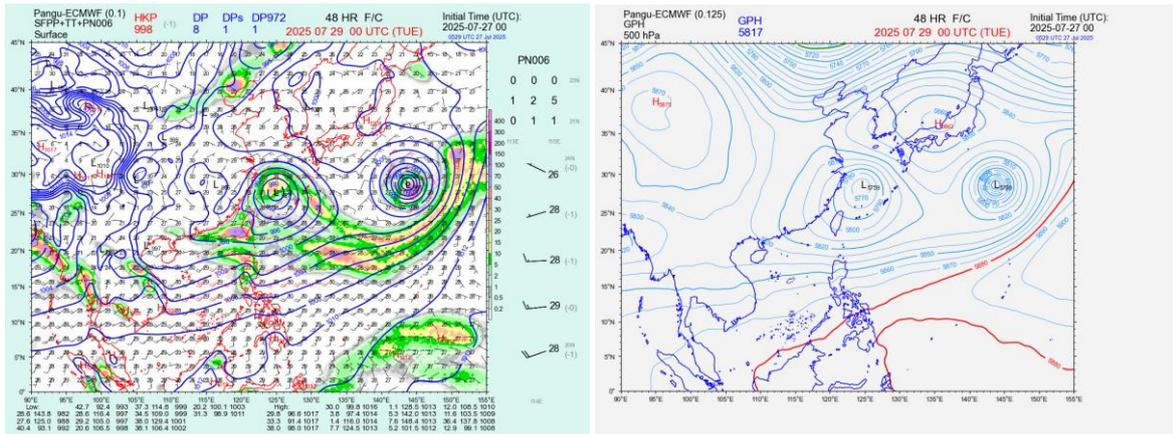

(a) and (b).

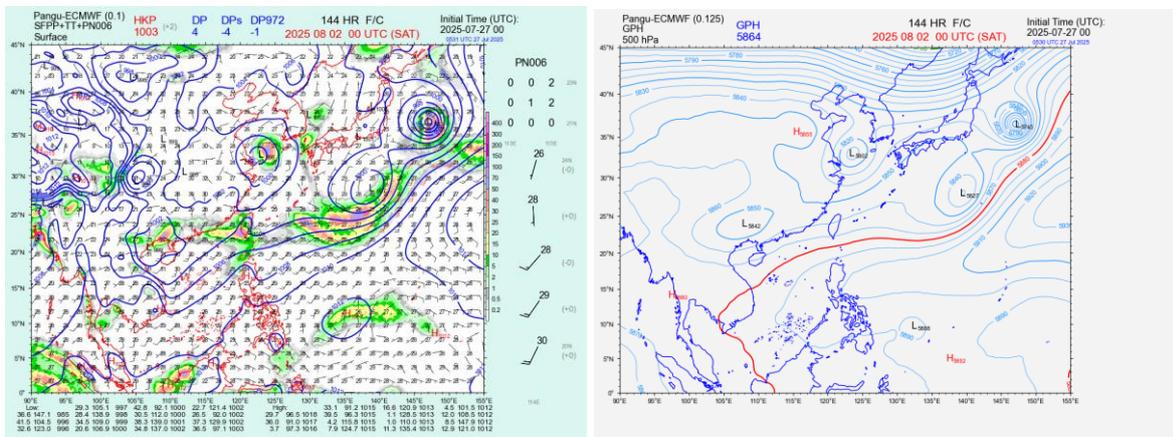

(c) (d)

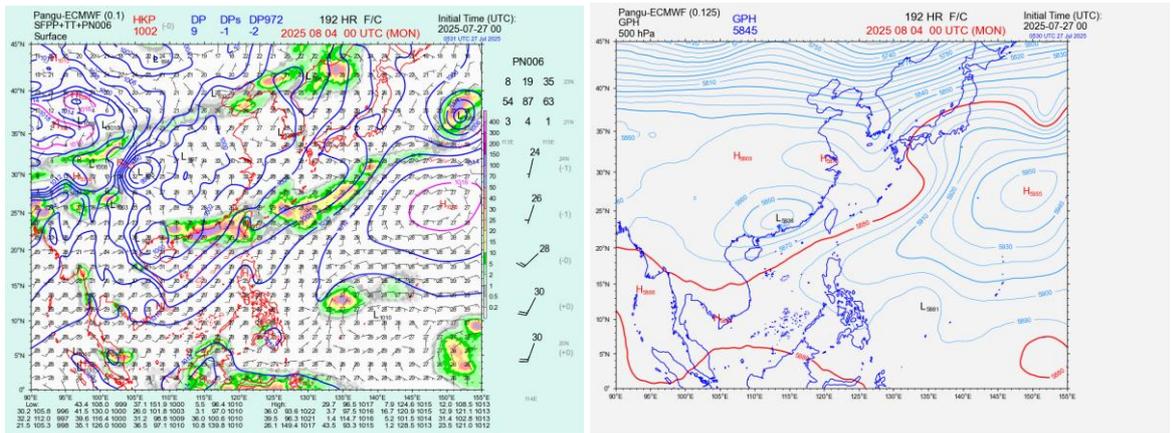

(e) (f)

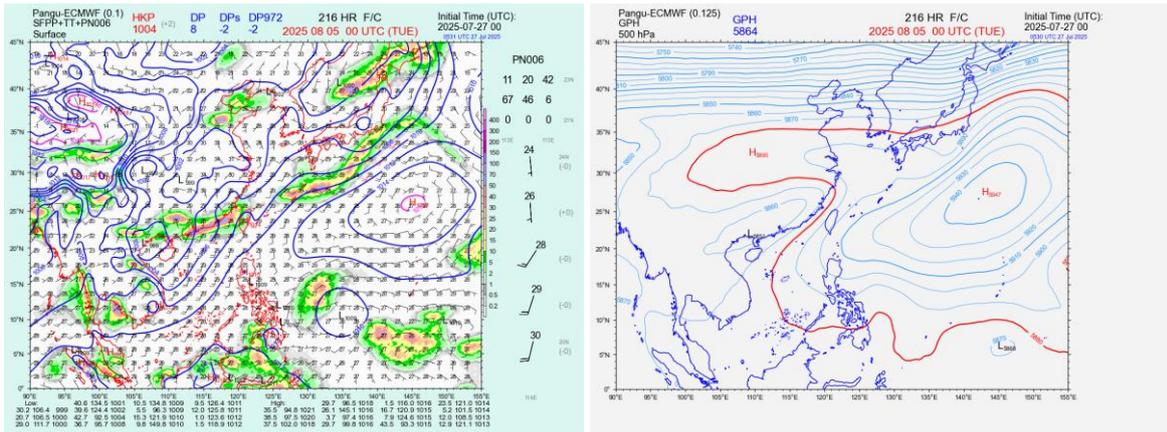

(g) and (h).

Figure 11 For the Pangu model run initialised at 00 UTC, 27 July 2025: (a), (c), (e), and (g) are the forecast surface wind field, surface isobar and the 6-h precipitation field; (b), (d), (f), and (h) are the forecast 500 hPa geopotential height field. The 5880 gpm curve is highlighted in red to represent the subtropical ridge.

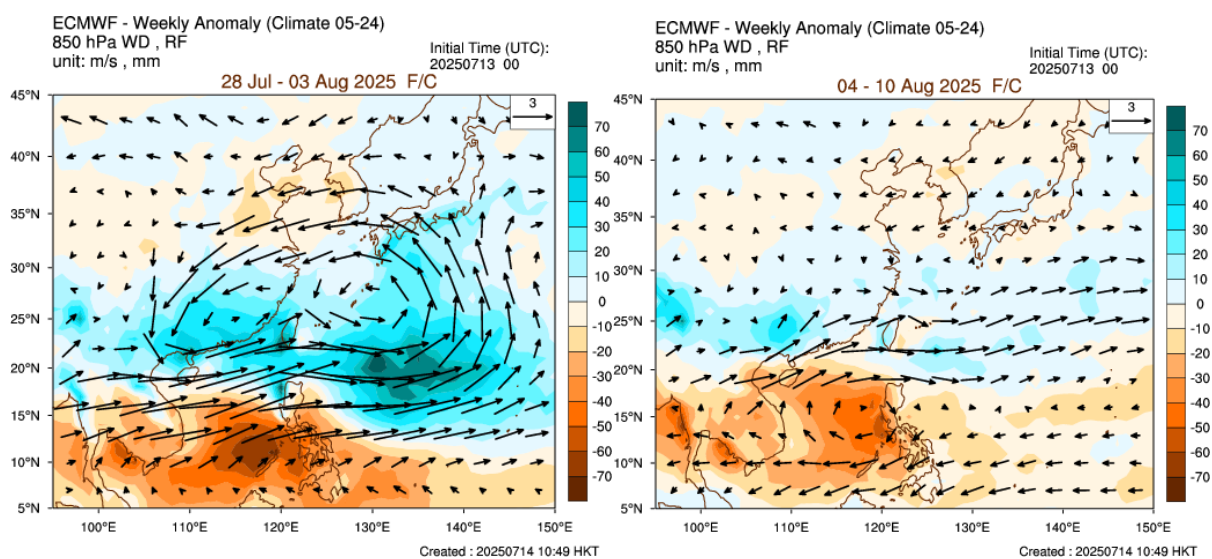

(a) and (b).

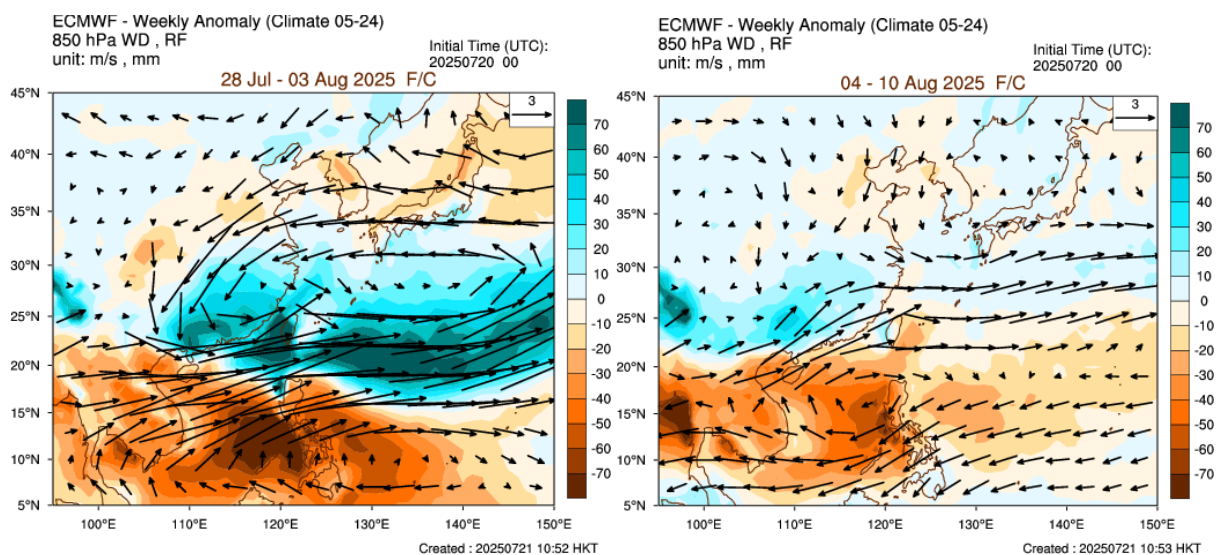

(c) (d)

Figure 12 (a) and (b) show the three-week and four-week sub-seasonal forecasts initialised at 00 UTC on 13 July 2025 showing the 850 hPa anomaly wind field and the surface precipitation anomaly field; (c) and (d) are the two-week and three-week forecasts initialised at 00 UTC on 20 July 2025.